# Living Without a Mobile Phone: An Autoethnography*

**Andrés Lucero**
Aalto University, Helsinki, Finland
lucero@acm.org

## ABSTRACT
This paper presents an autoethnography of my experiences living without a mobile phone. What started as an experiment motivated by a personal need to reduce stress, has resulted in two voluntary mobile phone breaks spread over nine years (i.e., 2002-2008 and 2014-2017). Conducting this autoethnography is the means to assess if the lack of having a phone has had any real impact in my life. Based on formative and summative analyses, four meaningful units or themes were identified (i.e., *social relationships, everyday work, research career,* and *location and security*), and judged using seven criteria for successful ethnography from existing literature. Furthermore, I discuss factors that allow me to make the choice of not having a mobile phone, as well as the relevance that the lessons gained from not having a mobile phone have on the lives of people who are *involuntarily disconnected* from communication infrastructures.

## Author Keywords
Autoethnography; autobiographical design; undesigning interaction.

## ACM Classification Keywords
H.5.m. Information interfaces and presentation (e.g., HCI): Miscellaneous.

## INTRODUCTION
On brink of burnout in late 2002, I decided it was time to get rid of the very tool that for the previous three years had allowed me to juggle with four simultaneous jobs as a web designer, a university lecturer, a professional soccer referee, and a freelance designer: my mobile phone. The idea of living without a mobile phone addressed a personal need of improving my life by exploring ways to reduce stress. Getting rid of my phone was neither intended as a research project [43], nor motivated by *"getting research points for it"* [27]. What started as a personal experiment, resulted in two voluntary mobile phone breaks (i.e., 2002-2008 and 2014-2017). Conducting this autoethnography is the means to assess if the lack of having a phone has had any real impact in my life.

I present an autoethnography [6,10] where I share my experiences living without a mobile phone over nine years, thus providing a long-term perspective on technology detox [23]. After a formative analysis based on a *retrospective account* [8] from the first period (i.e., 2002-2008), plus *reflections-in-action* [8], and *fieldnotes* [38] generated during the second period (i.e., 2014-2017), a summative analysis [8] was conducted where an overarching process of categorization and theming [38] took place, resulting in four meaningful units [8] or themes: *social relationships, everyday work, research career,* and *location and security*. Based on Duncan [8], Ellis et al. [10], and Schultze [38], I identify seven criteria for successful ethnography: *study boundaries, authenticity, plausibility, criticality, self-revealing writing, interlacing actual ethnographic material and confessional writing,* and *generalizability*. I discuss and use these criteria to judge this work. Factors that allow me to make the choice of not having a mobile phone and the relevance of these findings for people who are *involuntarily disconnected* from communication infrastructures that are increasingly taken for granted are also discussed.

The rest of this paper is structured as follows. I begin by reviewing the relevant related work. Then, I describe the case of the different 'no mobile phone' periods, followed by methods. Finally, I share my experiences living without a mobile phone, followed by discussion and conclusions.

## RELATED WORK
In this section, I briefly review two areas of related work that this paper builds upon. The first is an overview of autoethnography as a form of qualitative research, its history, and how autoethnography is done and judged. The second covers different uses of autoethnography in HCI (i.e., *traditional autoethnographies, autoethnographic approaches,* and *autobiographical design*).

### Autoethnography
Autoethnography [6,10] or personal ethnography [4] is a qualitative research form, an approach to research and writing that aims to describe and systematically analyze (*graphō* in Ancient Greek, 'writing') personal experience (*autós* in Ancient Greek, 'self') to understand cultural experience (*éthnos* in Ancient Greek, 'nation' or 'culture') [10]. Autoethnography fits into the tradition of confessional tales in ethnography [42] in which the researcher, who is repositioned as an object of inquiry [4,41], writes detail-rich stories from an emotional perspective [9,10] to depict a particular socio-cultural setting in terms of personal awareness and experience [4]. Autoethnography consists of well-crafted writing that can be respected by critics of literature and social scientists alike and must be emotionally engaging as well as critically self-reflexive of one's sociopolitical interactivity [41]. Bad autoethnography can be criticized for embodying the worst excesses of post-modernism [5], as the author creates a too self-indulgent [8,19,38], narcissistic [10,19,41], and individualized [5,8,19] narrative. Good autoethnography shares voices that might not have been heard [5,41], and insights that might have been too subtle to elicit [5,8,10].

*History of Autoethnography*

Denzin and Lincoln [6] identify six key 'moments' in the history of qualitative research. First, in the traditional period starting in the early 1900s, qualitative researchers aimed to create objective accounts of field experience. Ethnographers were then focused on exploring and describing the lives of *primitive* people, eager to show what life was like from the *native* point of view [8]. In the second modernist phase that spans from the postwar to the 1970s, qualitative research was concerned with reaching similar levels of rigor as its quantitative counterpart. The third moment between 1970 and 1986 was concerned with the blurring of genres.

The fourth moment starting in the mid 1980s is characterized by crises of representation and legitimation. Social scientists became increasingly troubled by social science's ontological (i.e., concepts), epistemological (i.e., knowledge), and axiological (i.e., value) limitations [10]. Scholars exposed the strong tie between the *facts* and *truths* that scientists *found,* and the vocabularies and paradigms the scientists used to represent them. *"There was an increasing need to resist colonialist, sterile research impulses of authoritatively entering a culture, exploiting cultural members, and then recklessly leaving to write about the culture for monetary and/or professional gain, while disregarding relational ties to cultural members."* [10] Scholars began wondering what would become of social sciences if they were closer to literature than physics, offered stories instead of theories, and were self-consciously value centered rather than pretending to be value free. Autoethnography will become a response to some of these dilemmas of social sciences in general, and ethnographic inquiry in particular [4].

In the fifth moment starting in the late 20th century, experimental writing and participatory research feature strongly. Narrative approaches typical of ethnography change to facilitate a more personal point of view by emphasizing reflexivity and personal voice [8]. Autoethnography emerges as a radical reaction to realist agendas in ethnography and sociology that privilege researcher over subject, method over subject matter, and maintain commitments to outmoded conceptions of validity, truth, and generalizability [6,41]. A researcher's decisions on who, what, when, where, and how to research, are necessarily tied to institutional requirements, resources, and personal circumstance [10]. Autoethnography acknowledges and accommodates subjectivity, emotionality, and the researcher's influence on research. Furthermore, researchers possess different assumptions about the world (e.g., speaking, writing, valuing, believing), differences that stem from race, gender, sexuality, age, ability, class, education, or religion. Canonical ways of doing and writing research advocate for a white, masculine, heterosexual, middle/upper-classed, Christian, able-bodied perspective, disregarding other ways of knowing, and rendering them as unsatisfactory and invalid [10]. Finally, in the sixth (postexperimental) and seventh (future) moments [6], fictional ethnographies and ethnographic poetry become taken for granted [19].

*Doing Autoethnography*

Using tenets of *autobiography* and *ethnography*, autoethnographers use hindsight to retrospectively and selectively write about past experiences (i.e., autobiography) that stem from studying (or being part of) a particular culture (i.e., ethnography), and/or possessing a particular cultural identity [10]. Social science publishing conventions require autoethnographers to analyze these experiences or *epiphanies* (i.e., remembered moments that have significantly impacted a person's life) using theoretical or methodological tools, and research literature [10]. Autoethnographers produce aesthetic and evocative thick descriptions of personal and interpersonal experience by identifying patterns of cultural experience based on field notes, interviews and/or artifacts, thus making characteristics of a culture familiar for insiders (i.e., cultural members) and outsiders (i.e., cultural strangers) [10]. Autoethnographers use storytelling, showing and telling, and alterations of authorial voice to produce accessible texts that describe these patterns, with the aim to reach a more diverse mass audience than the traditional research readership. Usually written in first person [5,8,10,19,38,41], autoethnographies can sometimes be expressed as a conversation between the author and the reader with the aim to offer lessons for further dialogue [10,38] (by posing new questions to be addressed by future research.)

*Judging Autoethnography*

Traditional criteria used to judge qualitative research may not be appropriate for autoethnography [12]. Acknowledging this, Duncan [8], Ellis et al. [10], and Schultze [38] have delineated key legitimacy and representation issues of autoethnographic accounts. Based on these issues, I have identified seven main criteria for a successful autoethnography. First, *study boundaries* [8] requires autoethnographers to describe the limits of their study using the four facets of time, location, project type and point of view. Second, *authenticity* [38] (also expressed as *reliability* [8,10] and *(construct) validity* [8,10]) refers to establishing a study protocol that would allow someone else to follow the researcher's procedures. Third, *plausibility* [38] (or *scholarship* [8]) relates to structuring the narrative according to the academic article genre and finding gaps in the research literature. Fourth, *criticality* [38] (referred to as *instrumental utility* [8]) entails guiding readers through imagining ways of thinking and acting differently. Fifth, *self-revealing writing* [38] consists of revealing unflattering details about the autoethnographer. Sixth, *interlacing actual ethnographic material and confessional content* [38] suggests that personal material be limited to relevant information in relation to the research subject. Finally, *generalizability* [10] (also expressed as *external validity* [8]) focuses on the readers who determine if the story speaks to them about their life or that of others they know.

**Autoethnography in HCI**
Though controversial due to HCI's epistemological goal of basing research on objective, third-party knowledge, different uses of autoethnography have begun to emerge in HCI: *traditional autoethnographies, autoethnographic approaches,* and *autobiographical design*.

*Traditional autoethnographies* in HCI that are written in first-person and systematically analyze personal experience to understand cultural experience are hard to come by. Instead, researchers tend to fall back to more traditional HCI practices when writing their autoethnographies, either by adopting a fully 'scientific' prose that confines the use of first person [20,31,35], and/or by concluding the autoethnography with a specific *design guidelines* section, or a concrete set of opportunities for design [2,20,35]. Two notable exceptions to this include Sengers's reflections on IT and pace of life [40], and Williams's use of personal fitness and self-tracking technologies to lose weight [43]. Through her highly personal stories about becoming aware of being oriented to time and work, Sengers aims to raise more general questions about the experiences of time and work that are tied to being modern as Westerners (and more specifically in her case as Americans). In a similar vein, Williams provides an intimate first-person account of the contradiction of being a researcher once critical of the logic behind diet control systems, and his use of them to lose weight, discussing the frequent negotiating of perspectives and boundaries from living with these technologies.

*Autoethnographic approaches* have been taken and applied by HCI researchers at different stages of the design process. Such approaches include using autoethnography as a first step when conducting user research [5], to inform the design of user studies [31], and as a lightweight method throughout an iterative design cycle [35]. Examples include Cunningham's MP3 player design student project [5], Efimova's reconstruction of personal blogging practices [9], Höök's accounts of horseback riding [20], Pijnappel's experiences with feedback systems to attempt skateboarding tricks [35], and O'Kane's use of a medical device during non-routine times [31].

*Autobiographical design* [27] focuses on design research that draws on extensive, genuine usage by those creating or building a system. Neustaedter and Sengers conducted expert interviews with 11 established HCI researchers to study actual examples of designing through self-usage. Prominent examples of autobiographical design include Erickson's Proteus [11] and Gaver's Video Window [13]. Erickson presents a subjective account of his experience with the design and long-term use (i.e., three years) of Proteus, a personal electronic notebook. While Erickson is aware of the objections to such an approach, which include concerns about subjectivity and lack of generalizability of the results, he argues that reflective analysis can yield much value if carried out carefully [11]. Gaver reflects on his (family's) particular experience of designing and living with the Video Window, an ambient weather display created over a number of months for his bedroom wall. Gaver argues that people can draw their own lessons from his particular experience with the Video Window, and thus autobiographical design can provide worthy lessons not focused on generalizability (as discussed in [27]).

**CASE**
In this section, I provide an overview (Figure 1) of the case on which this autoethnography is based, namely, the two periods of time when I went about living without a mobile phone. These voluntary breaks span nine years (i.e., 2002-2008 and 2014-2017), thus providing a long-term perspective on technology detox [23].

My work situation between 1999 and 2002 in Chile (Figure 1a) was that I had four jobs (Figure 1b). My day job was as a designer creating websites in a small 10-person company that also offered web hosting. Most of my activities here involved close face-to-face communication with my colleagues but would occasionally require that I were available to make quick updates outside office hours, which at the time were usually between 8:30 and 18:30. On Mondays and Wednesdays, I would also co-teach a third-year Bachelor course in graphic design between 17:00 and 20:00, with a half-hour subway commute to the university. On Tuesdays and Thursdays, I had training sessions between 18:30 and 20:00 as a professional soccer referee, with another half-hour commute by car to a different part of town. My refereeing duties also meant that on Saturdays and Sundays I would have to officiate between one and four soccer matches as a referee or assistant referee within a 500km radius from the city of Santiago, where I lived. Traveling to these games could sometimes start late on a Friday with an overnight bus trip to the city where the game would be played, and end by returning directly to work on Monday morning. Finally, I worked together with a business partner on extensive (e.g., six months) corporate identity projects as freelance designer, albeit only one project at a time.

The main use of my mobile phones at the time, first a Nokia 5110 and later a Nokia 3310 (Figure 1c), was to make and receive calls (no SMSs). My mobile phone allowed me to juggle with these four jobs and to take *ad hoc* requests, especially while commuting between jobs.

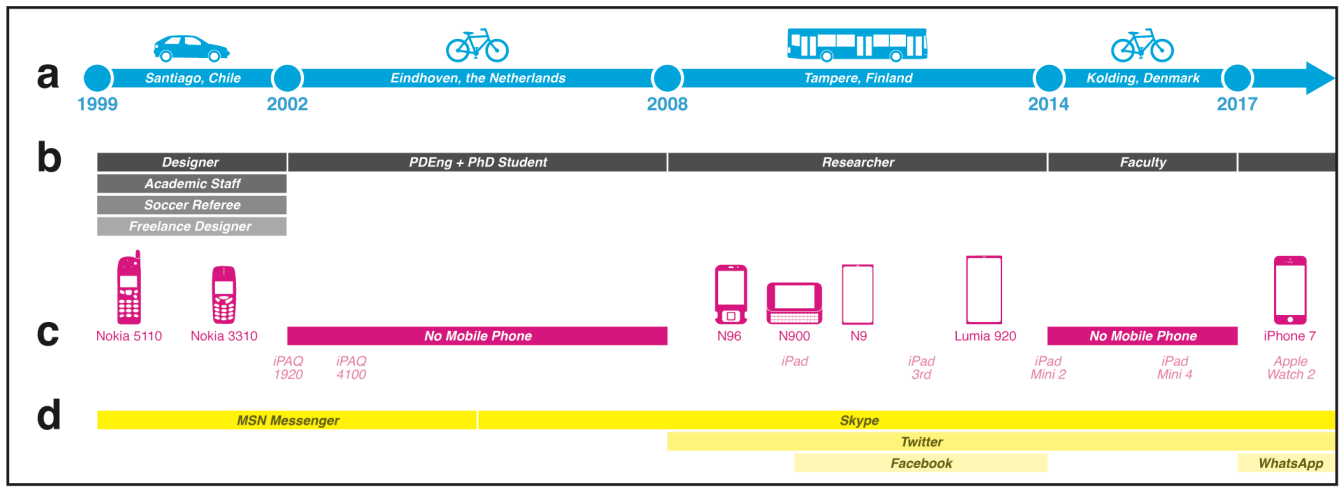

**Figure 1.** Timeline with an overview of: a) cities and countries where I have lived and main means of transportation used (cyan), b) main occupations (grey), c) 'no mobile phone' periods (2002-2008 and 2014-2017) plus mobile devices used at other times (magenta), and d) main instant messaging, video chat, and social networking services used (yellow).

I had the flexibility to leave work early, but that came at the cost of being always reachable. My mobile phone easily rang every hour, and sometimes even every 15 minutes. My mobile phone allowed me to work more, but as a result of a hectic schedule, I was on the brink of burnout.

**First Period Without a Mobile Phone**
On September 18, 2002, I moved to the Netherlands (Figure 1a), together with my partner, to study a two-year post-Master course. What started as a short break from the hectic life I led in Chile, was soon extended by a further four years to obtain my PhD (in fact, 16 years later we have yet to return home). We lived near the city center of Eindhoven and I would thus commute to work by bike, roughly a 10-minute ride. Our two children were born here and their mother took care of them at home. One conscious decision when moving to the Netherlands was that I would have just one job (Figure 1b). Part of this slowing down process also meant that I could get rid of my mobile phone (Figure 1c).

At the time, MSN Messenger was gaining popularity (it would be later replaced by Skype) (Figure 1d), which meant that I could use a laptop to chat and have voice (and video) conversations with my friends and extended family in Chile. I had internet access both at work and home in Eindhoven, thus I could communicate with my partner using her laptop, e.g., if she wanted me to get groceries on my way home after work. I had a landline at home mostly for our ADSL internet connection, and that we occasionally used to call our parents back in Chile. In essence, I could not identify a justifiable reason to own a mobile phone other than the convenience of being able to communicate when mobile, which, given I used a bicycle, was unfeasible.

**Back to Using Mobile Phones**
On September 18, 2008, my family and I moved to Finland (Figure 1a) where I resumed my work in industry now as a senior mobile HCI researcher (Figure 1b). My roles included acting as a researcher, designer, and strategist, which involved daily face-to-face meetings. We lived on the opposite side of the city of Tampere from where my office was, thus my commute to work was a 40-minute bus ride. Our children soon started going to the nearby daycare and later walked to school on their own, allowing my partner to go back to work. Although I had only one job, this one required that I use a mobile device once again.

As I worked for Nokia, I was expected to carry and use one of the company's mobile phones, by now called smartphones. I used four main devices, i.e., N96, N900, N9 and Lumia 920 (Figure 1c). Around that time, I also started using Twitter (October 2008) and later Facebook (February 2010) (Figure 1d), which I would mostly access from my smartphone to keep me entertained during the long bus ride home. Having a smartphone also meant I could reply to email from the bus, and so I would compose long emails, even using the N96's standard 4x3 keypad, which used multi-tap to cycle through letters and thus enter text. I mostly used my smartphone on the go to take pictures, check my email, coordinate meetings, browse the web, update my social network statuses, and seldom to receive and reply to SMSs and calls. Once at home, I would do most of these activities (plus reading) on an iPad without cellular capability (April 2010) (Figure 1c), except for calls and SMSs. Being always connected through cellular data when mobile or Wi-Fi at work and home meant I was becoming a distracted smartphone addict [17] (e.g., I repeatedly found myself using the device more of the time and becoming increasingly attached to it, choosing to use it at any opportune moment), and thus I was ready to once again disconnect. Instead of designing an app that would remind me to slow down [16,28], I decided to get rid of the phone altogether.

**Second Period Without a Mobile Phone**
On August 1, 2014, my family and I moved again, this time to Denmark (Figure 1a), returning to academia as faculty member (Figure 1b). Similar to the life we had in the

Netherlands, we lived in the city center of Kolding near the university so I could bike (a seven-minute ride) or walk (a 15-minute stroll) to work every day, while our two children took a 45-minute bus to school every morning. Once again, I got rid of my mobile phone (Figure 1c).

The main difference with my first mobile phone break in 2002, is that on top of being unable to receive and place calls (and SMSs), I was also disconnected from both work and personal email updates, the web, and social network notifications while on the go. I deactivated my Facebook account (November 2014) (Figure 1d), but actively used Twitter (and still do), especially at academic conferences.

**METHOD**
Living without a mobile phone during the aforementioned two periods (i.e., 2002-2008 and 2014-2017) addressed a personal need of improving my life by exploring ways to reduce stress. It was only at the start of the second period, after conversations with colleagues and inspired by the likes of PSY during his honest, unassuming, and frank closing plenary at CHI 2015[1], that I began considering the idea of writing an autoethnography of my experiences living without a mobile phone. Thus, getting rid of my phone was neither intended as a research project [43], nor motivated by *"getting research points for it"* [27]. Conducting this autoethnography is the means to assess if the lack of having a phone has had any real impact in my life.

**Data Collection**
My first step was to develop a *retrospective account* [8] of my life without a mobile phone during the first period (i.e., 2002-2008). These retrospective accounts (or headnotes [32]) consisted of events, experiences, and interpretations that were constructed from memory [9], using projects, notebooks, photographs, and emails to aid recall [10]. I was familiar with retrospective accounts from my time assessing first- and third-year Bachelor students' self-reflections in the Netherlands. In an at-the-time novel competency-based education system, students took the role of 'junior employees' and as such were responsible for their own competency development, choosing their own learning path. My role back then was to help them plan, read, and give feedback on their competency development. Switching roles to write my own retrospective accounts allowed me to identify important themes in my daily life that helped refine this study's focus, guide the ongoing literature review, and develop a language of description for my reflections.

During the second period (i.e., 2014-2017), I collected *reflections-in-action* [8] consisting of biweekly handwritten and digital notes taken on a notebook or iPad, respectively. These reflections-in-action were complemented by emails, photographs, and tweets. In addition, whenever traveling I recorded *fieldnotes* [38], which I tried to write on the spot, or as soon as possible after the event.

---
[1] https://twitter.com/emax/status/591149505170382848

**Data Analysis**
After a formative analysis based on *retrospective accounts, reflections-in-action*, and *fieldnotes*, a summative analysis [8] was conducted where an overarching process of categorization and theming [38] took place. Recurring problems, changes in attitudes, and significant concerns emerged after deeper and more detailed reflections, which developed into meaningful units [8]. These units or themes form the foundation of this autoethnographic narrative.

In addition, I experimented with themes by drawing tables [38] and with different types of visualizations to help clarify my thinking and keep an overview, similar to the one shown on Figure 1. I also shared my experiences, my initial interpretations, and drafts at different stages of this autoethnography with colleagues and extended family members [10,31,38]. Doing so allowed me to gather new perspectives and offer alternative interpretations. Finally, I compared and contrasted my personal experience against existing research [10].

**LIVING WITHOUT A MOBILE PHONE**
In this section, I will share my experiences as a *mestizo* (i.e., Latin American of mixed race), male, heterosexual, middle-aged, middle class, postgraduate-level educated, atheist living without a mobile phone over nine years spread in two periods (i.e., 2002-2008 and 2014-2017) along four main themes: *social relationships, everyday work, research career,* and *location and security*.

**Social Relationships**
Within HCI, substantive concerns with the limitations and negative effects of technology have started to emerge. These concerns span a diverse range of social, environmental and moral issues (e.g., climate change and e-waste pollution, busyness and overwork [14,22,40]). Undesigning technology (or interaction) [34] is a theoretical framework for conceptualizing the *intentional and explicit negation of technology by design*. According to Pierce, there are two main motivations for undesigning [34]: things we might consider bad, evil, or unnecessary (e.g., nuclear power plants, software updates that run slower), and things we might consider good, useful, or necessary, but which might be even better in lesser forms (e.g., iPad games, 24/7 internet access). The spectrum of technology negation *"ranges from the inhibition of particular uses of technology, to the broader displacement of technology, to the total erasure of an existing technology or foreclosure of an emerging one."* *Inhibition* is design that prevents the use of technology at the level of individual interactions in particular ways or contexts (e.g., Drift Table [39], speed bumps, displaying environmental costs of electricity consumption). *Displacement* is design that more substantially hinders the use of technology at the level of routine social practices (e.g., encourage dwellers to move TV sets from certain areas of the home). *Erasure* is design that completely eliminates a technology at the societal or existential level. *Foreclosure* targets a technology that has not fully emerged into existence (e.g., toxic substances, genetically modified foods). Pierce's framework for (un)design suggests this paper's work is a *displacement of technology* (i.e., mobile phone), whereby the phone has been removed from daily use. Reflecting on my personal experiences living without a mobile phone will hopefully raise more general questions tied to being modern.

As the mobile phone does not play a central role in my life, I do not feel the need to tell people about my lack of phone upon first encountering them. People first hear about it when, in the middle of a face-to-face conversation, a need to share information or make an appointment emerges:

*"I'll text you."*

*"Do I have your number?"*

Initially, and for many years, I felt embarrassed to admit to not having a mobile phone. Instead, I would avoid confessing and suggest communicating via email, or agreeing to our next appointment on the spot. Lately, I have opened up to explaining why I will not be able to receive a text message or call back. Some people react by joking about it:

*"How modern of you!"*

*"You're an anarchist."*

Others are puzzled and start asking questions, first in relation to urgent matters that normally pertain to the domain of the phone:

*"How do you do it with the kids (in case of an emergency)?"*

*"How do you communicate with your partner during the day (in case she needs to talk to you)?"*

A few start making assumptions, thinking I will not be reachable to pick up something on my way from the office to home to get some missing ingredients for the night's dinner, or if the menu has changed altogether:

*"So your partner gets to do all the groceries!"*

Then people's questions shift to the domain of a portable computer or smartphone:

*"How do you check or respond to email?"*

*"Are you on social networks?"*

With every question, people share what is truly important to them: being reachable to and able to reach their contacts; being able to multitask and handle basic work-related tasks on the go (e.g., email, calendar); being able to stay in touch with their loved ones, wherever they may be in the world. These assumptions work on the premise that people want to be able to reach and be reached by others anytime, anywhere, on the spot. Although there are a few mobile phone owners who do not tend to all communication systems all the time, this is probably the reality for most people. Peters and ben Allouch discuss the benefits of being *always connected* [33], accessible at all times and places, and how it makes technology-enabled people almost automatically adapt the new mobile communication technology into their daily life. I, on the other hand, was somehow exposed (or forced) to deal with disconnection at an early age.

In 1984 at the age of nine, my parents, my siblings and I moved to Switzerland, leaving my grandparents behind in Chile. At the time, due to cost, distance and time, going back to visit my grandparents was something that could only be done once every two years. In between visits, writing and sending letters was the only way to stay in touch with my grandparents (the cost of international landline phone calls was prohibitive). A letter would roughly take one month to travel between countries. While the wait was excruciatingly long for a nine-year-old, it also created a sense of expectation. Perhaps, I somehow got used to waiting for communications to come through (i.e., letters from my grandparents) already as a child, which resonates on how I deal with disconnection (by not having a mobile phone) as an adult. Today, people can immediately communicate with another person by calling them, sending them a text message, or starting a video call using Skype.

My children are now living a similar situation as I did when I was nine. A few years ago, we celebrated Christmas by having my parents and my in-laws set on separate laptops in different parts of the room. My daughter seamlessly transitioned from showing to us, to her grandparents in Santiago, Chile, and to her other grandparents in San Felipe, Chile what Santa had brought her. As far as my daughter was concerned, all six adults were present in the room. In a study of everyday dwelling with WhatsApp, O'Hara et al. [30] found that *'being together does not equate to the people in question sharing the same physical space; the shared domain is at once real and virtual but tied together into a delicate world of felt-life.'*

Certainly, there are factors that allow me to make the choice of not having a mobile phone. The fact that my extended family and the majority of my friends are far away living in another continent means that they are reachable through email and Skype (as depicted in the previous story), so I have no need for a mobile phone. I also do not feel cut off from the local community and other parents at school as those interactions happen mostly face-to-face or via email. However, if I lived in Chile, it would certainly be more challenging to keep in touch without a mobile phone.

Another factor is that my immediate family has been extremely supportive. My partner has had her own mobile phone throughout this time period (i.e., 2002-2017). During my two periods without a mobile phone while living in the Netherlands and Denmark, I could communicate with her via MSN and Skype, respectively. When our two children took a bus to school every morning in Denmark, we benefitted from the fact that in Finland, when children turn seven years of age, they get their first mobile phone together with the keys to their family's home so they can go to school on their own [25]. This means that my daughter also had a mobile phone in Denmark to communicate with her mom and let her know that she and her brother had reached school safely. If I was a single parent, I could not be constantly disconnected.

**Everyday Work**

Mobile devices are inherently tied to our identity and our experiences of being modern [40]. However, due to the role that smartphones have in the pervasive culture of busyness [22] in everyday life, different academic studies and autoethnographic-styled accounts[2,3] have begun uncovering some of the issues behind our *constant connection* [1,26,33], and *total availability* [26]. In response to this, communities of *techno-resistance* [15,44] and *technology non-use* [3,21,36], individuals actively setting boundaries and disconnecting from other devices [1], plus notions of *slow design* [16,28,29] and *digital detox* [23] have emerged.

At the university campus in Kolding, life revolves around teaching in lecture rooms, doing research at the lab just off campus, doing administration tasks behind my desk, visiting the campus library and cafeteria, and thus implies some degree of mobility. Naturally, people's frequent transitions within campus pose challenges when trying to find each other. I try to make myself available to colleagues and students who know they can find me most of the time between 9:00 and 16:00 (regular working hours in Denmark) at my desk in a shared open office. My social interactions with colleagues and students typically happen in a collocated fashion, thus I do not feel particularly isolated from them. Most people also know I do not have a mobile phone, so they are aware that calling me or responding to email on the spot when I do not have my laptop open is simply not possible for me.

I try to respond to email when I can, which can be on the spot but more frequently is within the first couple of hours, and at the latest by the end of the day. As a result of my ongoing attempt to reduce stress, my notifications are thus almost non-existent. However, being unable to take in calls and respond to emails on the spot makes me more prone to receiving so called *urgent* emails. For a recent trip to the US, I needed to spend one night in San Francisco. I asked the university's travel agency via email to please help me find a place downtown for a reasonable price. The next day, I received the following email:

*"Could you please call me. URGENT."*

Different aspects of the email indicated that the sender requested a prompt response: asking for a phone call instead of an email response; the use of the word 'urgent'; the use of caps lock. I swiftly called back using my partner's phone. It turned out the agent had found a hotel that had one last room available for me for a convenient price, hence the urgency. I confirmed the hotel room and thanked the person for their diligence. As I hung up the phone, I kept wondering about the actual urgency of the message. Over time, I have learnt that *urgent* emails are not always urgent. Such is the amount of emails that people receive on a daily basis [18], that I ask myself how my colleagues deal with their inbox. A few of them have resorted to simply ignoring certain emails:

*"If it's important, then they'll send it again."*

Perhaps aware of this, people also flag emails as *high priority*, hoping that a particular email will not be ignored. But whose expectations are we fulfilling by being available 24/7 for work-related issues? I have not yet signed a contract that would require me to be on call. Unless one is a rescue worker (e.g., doctor, a policeman or firefighter) or another occupation where timing is key (e.g., stock broker), then the odds are one can probably delay a response and people would barely notice. In fact, there are apps for this[4].

In relation to my work, being an academic in a Nordic country means that people respect the delay in my responses to email communications. Again, if I were in other cultures such as that of the US or Chile where it is important to be busy [23], then this would be less acceptable. Being an expat also means that I can get away with behaving in a slightly different way than the local Danes (e.g., not having a mobile phone, not responding to emails on the spot).

**Research Career**

Only recently at various social events during the CHI conference, I have opened up about the fact that, at different moments in my life, I have not had a mobile phone for extended periods of time. As the levels of mobile phone ownership have continued to rise, and the evolution of society to one where it is seemingly 'normal' to have a mobile phone, I occasionally felt some sort of peer pressure to own a mobile phone.

In 2008, when I was completing my PhD in Eindhoven, I attended the MobileHCI conference for the first time. As the conference was in Amsterdam and I was about to join Nokia in Finland, it only made sense that I attend. During a keynote, one speaker said:

*"I assume everyone here has a mobile phone in their pocket. Who doesn't have a mobile phone?"*

Thinking there would be more people in the room without a mobile phone, I put my hand up. As I was sitting at the front of the auditorium, I could not see other people's hands up in the air. Then I turned around to realize I was the only

---

[2] @baratunde (2013). Baratunde Thurston left the Internet for 25 days, and you should, too.

[3] @TomGoulding (2017). How getting rid of my smartphone revolutionised my life.

[4] Mailbox app. https://www.mailboxapp.com

one with their hand up. As expected, people's reaction was very respectful; nobody laughed and there was no communal gasp. But that's as far as empathy went. People attending the keynote that day were probably in shock. I remember vividly a couple conference attendees greeting me during the coffee break with the following:

*"You're the guy without a phone!"*

I draw upon this situation when I ask my students the same question at the start of my lectures on mobile HCI-related topics. I try to empathize with those who put their hands up by saying:

*"You're not the only one."*

People tend to assume that having a mobile phone is a requirement to doing research in (mobile) HCI. Whenever confronted with this, I tend to respond that people work on all sorts of research where they do not possess or have easy access to the technology that they are working on, e.g., augmented reality glasses or media façades. I also provide a personal example, whereby in my PhD research I worked on creativity support environments for industrial designers using tools such as interactive tables and walls. While I only had access to those at the lab, there are other ways to keep up to date with development in these areas, for instance, by attending the first ever Interactive Surfaces and Spaces (ISS[5]) conference in Australia (at the time called TableTop). Often (design) students claim that they do not want to do too much research on a given topic to keep a fresh mind [7]. I personally do not support this view, since one should make every possible effort to be familiar with the state-of-the-art in a given field. Having said that, owning a given gadget or piece of technology should not be a requirement for performing research in that particular field. Next, I will try to illustrate why.

When I first stopped using a mobile phone in 2002, phones were mostly used to make and receive calls. Then, when I joined Nokia in 2008 and had a mobile phone once again, these devices included new functionality (e.g., camera, web browser, apps, etc.). My first project was to work on a social network service built around sharing personal geotagged photos [24]. While I was familiar with camera phones and GPS, there were other things such as SMSs, MMSs, and SIM cards that simply were not yet in use in Chile in 2002. Being an outsider to the field allows applying frames of reference from other domains. In practice, this implied trying out what worked (and what did not) in my previous research (i.e., surface computing) to the field of mobile HCI. This sometimes led to naive remarks or ideas that were either already available in mobile phones or were simply not possible at the time. However, many times it resulted in making the research team approach a given design problem from a different perspective.

---

[5] Interactive Surfaces and Spaces. https://iss.acm.org/

Many times, my colleagues would have explored an idea years before and simply would have given up on it because the technology available at the time would not allow implementing it, or because they failed to see the potential. In that respect, my team was formed mostly by middle-aged male engineers (like me) that had been working for the company for an average of 12 years (roughly the same number of years I had been working on interaction design projects). There was a certain way of doing things that was established and ideas would be quickly discarded by saying:

*"We tried that years ago."*

*"It cannot be done."*

Initially, I would take their word for it, but as I gained more experience and started to see that this was a way of thinking that was perpetuated inside the company, I began to further push these *known ideas* and give them a real opportunity to develop them into projects before discarding them. The ability to remain oblivious to the current set of features offered by mobile devices meant that I was able to explore ideas around what could motivate (my) usage: what *should* mobile devices do and how *should* we interact with them. In a sense, it allowed me to focus on developing ideas and improvements for mobile devices that were revolutionary in nature, and not constrained by the status quo.

**Location and Security**

When traveling to another country for a conference or on holiday, my colleagues (and people in general) tend to rely on their mobile devices to figure out where they are and where they need to be. Recently, I was trying to give directions on how to get to my home to a visiting researcher from the UK that had just landed at our local airport. As I struggled to give him directions for the nearest bus stop by looking at a map and sending chat messages, he suddenly replied:

*"Got it! I know which bus to take and where to get off."*

Google had told him exactly what to do by fetching his current location and him providing the destination.

Without an internet connection abroad, I usually have to plan and visit the conference website beforehand and take a few screenshots that are stored in my tablet's photo library or download a PDF that I can view offline. For most cases, this strategy works fine. Sometimes, however, I will only remember roughly where the conference hotel is located and begin wandering the city. By consulting local maps on the street and asking passers-by for directions, I often find my destination. While in general people are willing to help a tourist or a visitor that seems lost or looking for an address (with or without an open map in their hands), it is becoming increasingly common for people to assume that, just as they are able to find their way around pretty much anywhere in the world with the help of their mobile devices, so should everyone else (that owns a mobile phone).

During a trip to San Francisco, I tried to book a transfer from SFO airport to the city of San Jose. I took out my laptop (I traveled without my non-cellular iPad Mini) and used JFK airport's 30-minute free Wi-Fi access to go online. The website[6] had a mobile phone number as a compulsory field (I gave my partner's number in Denmark) plus an option to indicate if one preferred to be reached by call or text. Even if one would have a mobile phone available, people living outside the US would probably want to avoid extra roaming charges, be it due to call or text messages. Most airports provide 30-minute or free Wi-Fi access these days, then why not default to email? Why make having a phone number compulsory? This issue is of course not only related to traveling but is more generally linked to the use of a mobile phone as a security measure.

In Europe, it is often common for companies and universities to provide corporate credit cards 'for approved business use only' to employees in senior positions and faculty who travel often. Recently, I was trying to register online for an academic conference. When it came to the point of processing the payment, the system tried to send a text message with a security code. Unable to receive such a message, a colleague kindly volunteered to let me provide her number and process the payment. A couple of months later on a Sunday night, I was in a similar situation trying to register for another conference. When I called my colleague to get the security code from her, she kindly (and rightly so) reminded me that I would have to change that number eventually. While this could be perceived as me simply leveraging other people's devices to accomplish my tasks, shifting my inconvenience of not having a phone to others on a regular basis, this has only happened a few times over my two periods without a mobile phone.

Another example where the mobile phone is used to send secure information as a text message sometimes occurs when setting a new email account. I received an email on my Gmail account informing me that my new university email account and password had been set up. The normal procedure required the person to visit IT Services in person with a photo ID in order to receive their credentials. As I was out of the country at the time, they requested a phone number to send the information securely, indicating that sending the credentials to an unknown Gmail address was considered risky. I replied by suggesting that they send the information to my then current official university email address as an alternative. Perhaps assuming that I did not want to share my mobile phone number with them, they kindly but more firmly asked for my number once again, to which I replied by confessing I did not have one. Finally, they gave up and sent the credentials to the 'risky' non-secure and unknown Gmail address (instead of opting for the slightly more secure option that implies the use of an official university email address). Similarly, Twitter requires a mobile phone number to create a new account, and Google uses a mobile phone as a security measure by sending a security code to a person's phone in order to give them access to some feature (e.g., account configurations).

Going back to my San Francisco trip, upon landing in SFO I took out my laptop and the airport shuttle service had sent me an email requiring clients to complete a Self Check-In to indicate when ready for pickup. SFO provided free unlimited Wi-Fi at the airport, which meant I could collect my luggage and complete the check-in process. The website showed me in real time guest and vehicle number, as well as my and the vehicle's current position in Google Maps. Unfortunately, it was drizzling at the time, so my screen and keyboard got wet. Despite the usefulness of the app, there was no major difference between the info provided by the app and the one I had gathered from the person at the information desk where shuttle services pick up clients:

*"One floor up, out door three, across the walkway, and you'll see the sign."*

Would the driver have found me had I not completed the self check-in?

The feeling of having no mobile phone but having a tablet (i.e., without cellular capability) is similar to that of traveling abroad and deciding not to use roaming: one depends on free Wi-Fi spots, without knowing when the next one will be available for use. While to some readers this may create some sort of anxiety, when this is the default mode, one tends to worry less and less about not being reachable.

**DISCUSSION**

**What Makes This Autoethnography**

I will use the aforementioned seven criteria for successful ethnography [8,10,38] to judge this work. First, I have clearly defined *study boundaries* by stating the duration of the study (i.e., 2002-2008 and 2014-2017), the countries where it was conducted, what the study was about (i.e., living without a mobile phone), and the author's point of view that was adopted. Second, regarding *authenticity*, in the method section I have described in detail the *retrospective accounts, reflections-in-action*, and *fieldnotes* generated, and the summative analysis that followed, allowing the reader to reconstruct the research process and assess interpretation. Third, concerning *plausibility* I have structured the narrative as an academic article, and found gaps in the research literature. Fourth, to achieve *criticality* I have written this research as an autoethnography, which invites writer and reader to reflect and imagine new ways of thinking and acting. Fifth, I have achieved *self-revealing writing* by revealing unflattering details about myself. Sixth, about *interlacing actual ethnographic material and confessional content* I have restricted my personal stories to information that pertains to the lack of mobile phone and how it has influenced my life. Finally, with regards to *generalizability,* the personal stories of this autoethnography should have spoken to the reader about their life or that of others they know. A similar long-term autoethnographic approach of another researcher without a mobile phone or semi-structured interviews could help address some of the inherent limitations of the method [37].

---
[6] SuperShuttle. http://www.supershuttle.com

**Involuntary Disconnection**
The HCI community would normally expect this autoethnography to conclude with a specific *design guidelines* section, or a concrete set of opportunities for design. In line with Sengers [40] and Williams [43], this *traditional autoethnography* deliberately skips this and instead concentrates on systematically analyzing personal experience to understand cultural experience. Here, I aim to go beyond my own personal experiences and consider what a life without a mobile phone tells us about other parts of our society.

Next, I will attempt to develop empathies and insights into the lives of people unlike me, a privileged member of a hyper-connected and technology-saturated society. The reflections and the lessons gained from not having a mobile phone are also relevant to the lives of people who are *involuntarily disconnected* from communication infrastructures that are increasingly taken for granted. For instance, there are places in the world where no mobile service can be found at all; cultures where there are no Wi-Fi hotspots or alternative mobile technologies to compensate for the lack of a phone. Moreover, there are families and entire communities that share a single phone, rather than each person in each household owning a mobile device. Likewise, there are people who cannot use Facebook or Skype because of technology challenges. Furthermore, there are people who voluntarily choose to turn off their mobile phone in the evenings, plus a growing community who attend retreats to get an *off the grid* experience. Despite much of the fabric of our culture being conducted through the updated-every-hour connectivity that is accessible through a smartphone, the examples above illustrate situations where such connectedness is either not possible or wanted. On the other hand, there are examples of first-world people for whom mobile phones are the most powerful computing devices they have, for whom a laptop at home or even a desk at work is but a luxury. For others, their mobile phone is the only piece of computing equipment upon which to interact with all the apps and websites that demand their existence (my parents being a recent example of this).

**Back to Using a Phone**
Since January 1, 2017, my family and I are back in Finland, and I am still faculty member. We live in the city of Espoo, and my job is split between the city of Espoo, with a 40-minute commute to work by bus, and the capital city Helsinki a further 20-minute bus ride. Our children once again walk to school every morning. A smartphone and a mobile cellular subscription (both paid by the university) are offered by default upon starting employment. I opted for an iPhone 7 (Figure 1c) to explore the possibilities for interaction behind 3D Touch[8], and purchased an Apple Watch 2 (Figure 1c) to experience first-hand what constant notifications, activity tracking, and health tools could offer me. In addition, I began experimenting with lecturing by connecting my phone to the projector and using the watch to control slides. Initially, I felt self-conscious about raising my wrist to skip to the next bullet and slide, but audiences seem to feel it looks natural, and thus I have adopted this as my default way to present slides. Finally, I have started playing PokémonGo[9] both on my phone and watch.

As you can see, I have once again fully embraced life with a mobile phone, and I have even added a wearable to the mix. I am still on Twitter and began using WhatsApp (Figure 1d) but have found it slightly difficult to adapt to it simultaneously being a synchronous and asynchronous communication channel. I created a group to keep in touch with my extended family members. My parents and brother live in Chile and my sister lives in France, with a six- and one-hour difference with me, respectively. Being spread across time zones, it was not uncommon for my phone to start vibrating at odd hours in the middle of the night. While they seemed to be used to WhatsApp being an asynchronous medium, I on the other hand perceived it as a synchronous one. My extended family were kind enough to introduce me to the 'Do Not Disturb' feature on the phone, which is currently the default mode on my phone.

**CONCLUSION AND FURTHER RESEARCH**
In this paper, I have presented an autoethnography where I shared my experiences living without a mobile phone over nine years spread in two periods (i.e., 2002-2008 and 2014-2017). Based on formative and summative analyses, four meaningful units or themes were identified (i.e., *social relationships, everyday work, research career,* and *location and security),* which were subsequently judged using seven criteria for successful ethnography (i.e., *study boundaries, authenticity, plausibility, criticality, self-revealing writing, interlacing actual ethnographic material and confessional writing,* and *generalizability*). Factors that allow me to make the choice of not having a mobile phone and the relevance of these findings for people who are *involuntarily disconnected* from communication infrastructures that are increasingly taken for granted are also discussed. I hope this narrative will inspire other researchers to reflect on the larger role that technology plays in their lives.

**ACKNOWLEDGMENTS**
I would like to thank the anonymous reviewers for their helpful comments. Thanks also go to Martin Porcheron, Jesper Kjeldskov, Shaun Lawson, Soledad Paz, Carmen Vera, and Rayen Lucero for critically reading earlier drafts of this paper.

---

[8] 3D Touch. https://developer.apple.com/ios/3d-touch/

[9] PokémonGo. http://www.pokemongo.com